\documentclass{article}

\usepackage{epsfig,amsmath,amssymb,amscd,amsfonts,bm,subfigure,color,setspace,graphicx}

\usepackage{amssymb}

\newcommand{\ds}{\displaystyle}
\newcommand{\beq}{\begin{eqnarray}}
\newcommand{\eeq}{\end{eqnarray}}
\newcommand{\beqq}{\begin{eqnarray*}}
\newcommand{\eeqq}{\end{eqnarray*}}

\newcommand{\mn}{(m_1,\dots, m_N)}

\newcommand{\ANK}{A_{N,K}}
\begin{document}




\title{Coagulation-fragmentation for a finite number of particles and application to telomere clustering
in the yeast nucleus}
\maketitle
\author{Nathanael Hoze}\footnote{Ecole Normale Sup\'erieure, Institute of Biology (IBENS), Group of Computational Biology and Applied Mathematics, 46 rue d'Ulm 75005 Paris, France.}
\author{David Holcman$^{1,}$}\footnote{Department of Applied Mathematics, UMR 7598 Universit\'e Pierre et Marie Curie 187, 75252 Paris 75005 France.}


\begin{abstract}
We develop a coagulation-fragmentation model to study a system
composed of a small number of stochastic objects moving in a
confined domain, that can aggregate upon binding to form local
clusters of arbitrary sizes. A cluster can also dissociate into two
subclusters with a uniform probability. To study the statistics of
clusters, we combine a Markov chain analysis with a partition number
approach. Interestingly, we obtain explicit formulas for the size
and the number of clusters in terms of hypergeometric functions.
Finally, we apply our analysis to study the statistical physics of
telomeres (ends of chromosomes) clustering in the yeast nucleus and
show that the diffusion-coagulation-fragmentation process can
predict the organization of telomeres.
\end{abstract}
%
%
%
%



\section{Introduction}
Coagulation-fragmentation modelings have been applied to various
complex systems evolving at various scales ranging from star
formations to polymer organization. Although coagulation of a large
number of particles is described by continuum variable, for a system
composed of a finite number of stochastic particles,
Marcus-Lushnikov process \cite{Lushnikov} can be used. This approach
is based on Markov processes and combinatorial stochastic
processes \cite{Pitman,Smoluchowski,Lebowitz,Doering,Wattis}.
\\
In this letter, we study the dynamics of finite number of particles
undergoing coagulation-fragmentation in a confined domain. Two
stochastic particles bind with a Poissonian rate $k_f$ while a
cluster can coalesce with any other cluster or particle at the same
rate $k_f$, or dissociate at a rate $(n-1)k_b$, where $n$ is the
number of particles in the cluster and $k_b$ is the backward rate
constant. The dissociation gives rise to two clusters of size $p$
and $n-p$, where the law of dissociation uniform.  We develop a
Markov analysis to compute the probability of a distribution of $K$
clusters and obtain the steady state distribution, the mean, the
variance, the size of clusters and the number of particles per
cluster. Finally, we apply the present model to the organization of
the 32 telomeres in the yeast nucleus. \\
\section{Analysis of cluster dynamics }
\begin{figure}[ht!]
\begin{center}
\includegraphics[scale=0.65]{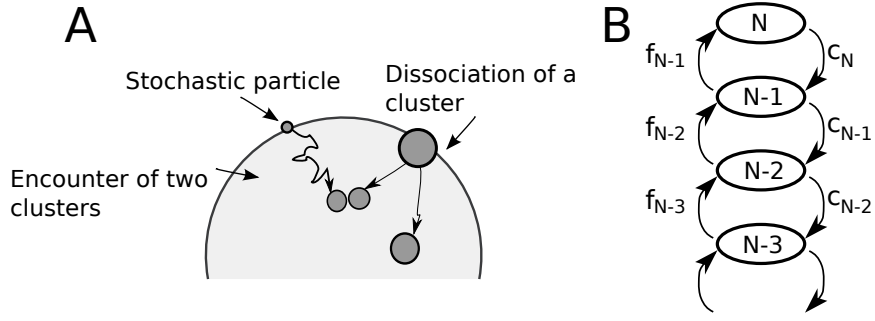}
\caption{(A) Coagulation-fragmentation of a finite number of particles. Two particles aggregate at a rate $k_f$ and a cluster of $n$ splits at a rate $(n-1)k_b$. (B) The Markov chain representation of the cluster dynamics where $f_K$ (resp. $c_K$) is the dissociation (resp. association) rate of a cluster (resp. of two clusters) when there are $K$ clusters.}
\label{fig1}
\end{center}
\end{figure}

We consider $N$ stochastic particles located in a confined domain
which can interact following the rules described above (Fig.
\ref{fig1}A). Our goal is to compute the probability density
function
\beq
P_{K}(t) =Pr\{ \hbox{ to have $K$ clusters }\}
\eeq
that the particles are distributed in $K$ clusters at time $t$. It
satisfies a Markov chain that we shall now derive: the probability
of having $K$ clusters at time $t+\Delta t$ is the sum of the
probability of starting at time $t$ with $K-1$ clusters and one of
them dissociates into two smaller ones plus the probability of
starting with $K+1$ clusters and two of them associate plus the
probability of starting with $K$ and nothing changes (Fig.
\ref{fig1}B). The first probability is the product of $P_{K-1}(t)$
by the transition rate $f_{K-1}\Delta t$ to go from the state where
there are $K-1$ clusters to $K$, while the second term describes the
transition from $K+1$ clusters to $K$. It is the product of
$P_{K+1}$ by the transition rate $c_{K+1}\Delta t$. For a set of $K$
indistinguishable clusters, the number of pairs is equal to
$\frac{K(K-1)}{2}$ and the association rate is
\beq
\label{eq:cK}
c_{K}= \frac{K(K-1)}{2}k_f,
\eeq
where $k_{f}$ is the encounter rate of two particles if there is no
other particle.  In contrast, for a cluster of size $n$, the
dissociation rate is $(n-1)k_b$. When there are $K$ clusters of size
$n_i$, the ensemble of configuration is $(n_1,\dots,n_K)$, with the
conservation of the number of particles $\sum_{i=1}^K n_i=N $. We
further considered that clusters are ordered by size $n_1
\geq \dots \geq n_K\geq 1$. The total transition rate from $K$ to $K+1$
clusters is the sum over all possible dissociation rates
 \beq
 \label{eq:fK}
 f_K = \sum_{i=1}^{K}(n_i-1)k_b= (N-K)k_b.
 \eeq
Thus the Master equation for $P_{K}(t)$ is
\cite{Redner2,Schuss:2010,Matkowsky}
\beq
\label{eq:PKt}
\left \lbrace
\begin{array}{ccc}
\dot{P}_1(t) &=&-f_1P_1(t)+c_2P_{2}(t)\\ &&\\
\dot{P}_K(t)
&=&-(c_{K}+f_K)P_K(t)  +c_{K+1}P_{K+1}(t)+f_{K-1}P_{K-1}(t)\\
&& \label{Markov1}\\
\dot{P}_N(t) &=&-c_NP_N(t)+f_{N-1}P_{N-1}(t).
\end{array}\right.
\eeq

\begin{figure}[ht!]
\begin{center}
\includegraphics[scale=0.6]{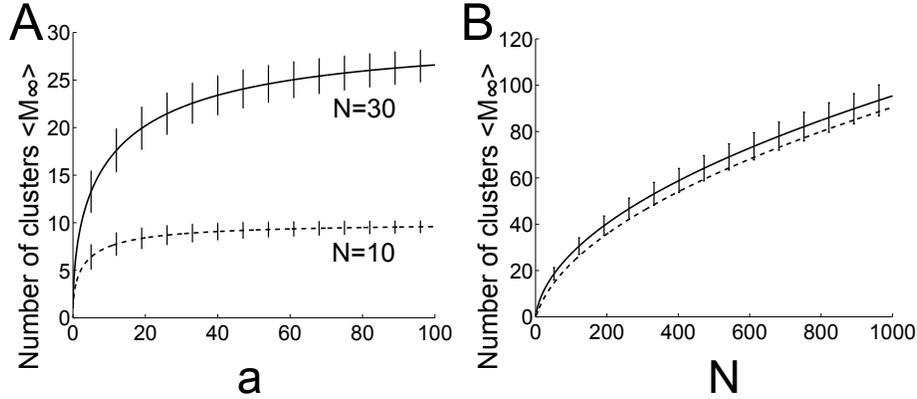}
\caption{ Number of clusters $<M_{\infty}(a,N)>$ as a function of the equilibrium parameter $a$ (A), for $N = 10$ and $N=30$ obtained from eq. \eqref{Minf} and (B) as a function of number $N$ for $a=5$, using eq. \eqref{Minf} (continuous line) and the asymptotic approximation eq. \eqref{asymM} (dashed line).}
\label{fignumber}
\end{center}
\end{figure}

To determine the number of clusters at steady state, we integrate
explicitly the Markov chain \eqref{Markov1} and the steady state
probability $\Pi_K= \lim_{t \rightarrow \infty} P_K(t)$ is given by
\beq
\Pi_{K+1}=\frac{(2a)^K}{K!}\frac{(N-1)!}{(K+1)!(N-K-1)!} \Pi_1,
\label{eq:Pik}
\eeq
where the equilibrium parameter is
\beq
a=\frac{k_b}{k_f}.
\eeq
Using the normalization condition $\sum_K\Pi_K=1$, the probability $\Pi_1$ can be expressed using hypergeometric series:
\beq
\Pi_1=\frac{1}{ _1F_1(-N+1;2;-2a)},
\eeq
where $_1F_1$ is the Kummer's confluent hypergeometric function (Eq. 13.1.2, \cite{AbraSteg}).
At steady state, the average number of clusters is
\beq
<M_{\infty}(a,N)>= \sum_{K=1}^N K\Pi_K =1+a(N-1)G_1(a,N)
\label{Minf}
\eeq
where
\beq
G_1(a,N)=\frac{_1F_1(-N+2;3;-2a)}{ _1F_1(-N+1;2;-2a)}.
\eeq
In addition, the $n$th-order moment $\mu_n$ is given by
\beq
\mu_n = \sum K^n\Pi_K=\frac{H^n ({_1F_1})(-N+1;2;z)_{|z=-2a}}{_1F_1(-N+1;2;-2a)},
\eeq
where $H$ is the operator defined by $H(f)(z) = \frac{d}{dz}zf(z)$.
Since the derivative of Kummer's function is
\beq
 \frac{d{_1F_1}(\alpha;\beta;z)}{dz}=\frac{\alpha}{\beta} {_1F_1}(\alpha+1;\beta+1;z),
\eeq
the moments can be written as
\beqq
 \mu_n&=& \sum_{k=0}^{n}\alpha_k^n\frac{(N-1)!}{(k+1)!(N-1-k)!} \frac{(2a)^k}{k!}G_k(a,N),\\
 &=&  \sum_{k=0}^{n}\alpha_k^n \frac{\Pi_{k+1}}{\Pi_1}G_k(a,N),
\eeqq
 where
\beq
G_n(a,N)=\frac{_1F_1(-N+1+n;2+n;-2a)}{ _1F_1(-N+1;2;-2a)}
\eeq
and the coefficients $\alpha_k^n$ are given by
\beq
\alpha_k^n = \left \lbrace
\begin{array}{ccc}
k!\sum_{j=0}^{k/2} (-1)^j\ds{\frac{(k+1-j)^n+(j+1)^n}{(k-j)!}} \mbox{ if $k$ is even,}\\
\\
 k!\sum_{j=0}^{(k-1)/2} (-1)^j\ds{\frac{(k+1-j)^n-(j+1)^n}{(k-j)!}} \mbox{ if $k$ is odd,}\\
\end{array}\right.
\eeq
and $\alpha_0^n= \alpha_n^n=1$. Thus the variance of the number of clusters is given by
\beq
<V_{\infty}(a,N)>&=& \mu_2-\mu_1^2\nonumber \\
&=& a(N-1)G_1(a,N)
+  \frac{2}{3}a^2(N-1)(N-2)G_2(a,N)
- a^2(N-1)^2 G_1^2(a,N).
\eeq
In addition, for large $N$ and fixed $a$, using asymptotic results for hypergeometric functions \cite{AbraSteg}, we have
 \beq
G_n(a,N)&\approx&   \frac{(n+1)!}{(2aN)^{n/2}},
\label{asymGn}
\eeq
and we obtain from equation \eqref{Minf} that the mean number of
clusters is
\beq
<M_{\infty}(a,N)>=\sqrt{2aN}+O(1).
\label{asymM}
\eeq
As a result of the above analysis, we can estimate the mean number of clusters $<M_{\infty}(a,N)>$, which is plotted as a function of $a$ (Fig. \ref{fignumber}A) and  $N$ (Fig. \ref{fignumber}B).
\section{Size of the clusters}
Because the previous analysis cannot predict the size of the clusters,  we shall now evaluate the probability density function for the number of clusters of different sizes. This analysis relies on carefully studying the configurations of $N$ particles decomposed in $K$ clusters. The configuration is described by the ensemble of ordered clusters of size $n_1,\dots,n_K$
\beqq
\ANK= \{(n_j)_{1\leq j \leq K} ; \sum_{i=1}^K n_i=N   \mbox{, } n_1 \geq \dots \geq n_K\geq 1\},
\eeqq
which is equivalent to the set
\beqq
\ANK'= \{ (m_i)_{1\leq i \leq N} ; \sum_{i=1}^N im_i=N \mbox{, } \sum_{i=1}^{N}
m_i=K\},
\eeqq
where $m_i$ is the number of clusters of size $i$. The
correspondence from $\ANK$ to $\ANK'$ is obtained by taking into
account that for all clusters $n_j$ of size $i$,  we have the
relation $m_i=\sum_j \mathbb{I}(n_j=i)$ \cite{Pitman}. We enumerate
the number of configurations associated with the ensemble $\ANK'$.
It is the same as counting the occurrence of the integer $i$ in the
decomposition of $N$ as a sum of $K$ integers. The number of
occurrence of the configuration $\mn$, when there are $m_1$ clusters
of size 1, $m_2$ clusters of size 2 ..., is the multinomial
coefficient $ \frac{K!}{m_1!\dots m_N!}$. Thus the conditional
probability of the $\mn$ distribution, when the total number of
clusters is equal to $K$, is obtained by normalizing the equilibrium
over all possibilities
\beq
\label{eq:fproba}
p(m_1,\dots,m_N|K)
=\frac{\ds{\frac{K!}{m_1!\dots m_N!}}}{\sum_{(m_i) \in \ANK'} \ds{\frac{K!}{m_1!\dots m_N!}}}.
\eeq
Interestingly by summing the series $(X+X^2+...+X^N)^K$ for the
$N^{th}$ order coefficient and using the $N-K$ derivative of
$\frac{1}{(1-X)^K}$, we have
\beq
\sum_{(m_i) \in \ANK'} \ds{\frac{K!}{m_1!\dots m_N!}} = \frac{(N-1)!}{(K-1)!(N-K)!}.
\label{explicit}
\eeq
At this stage, we shall explain the rational for expression
\eqref{eq:fproba}. Indeed, it corresponds to the equilibrium
distribution of particles in clusters, where the dissociation (resp.
association) rate is proportional to the number of elements (minus
one) (resp. the number of pairs of particles).

\begin{figure}[ht!]
\begin{center}
\includegraphics[scale=0.5]{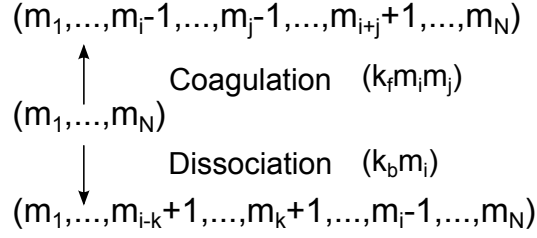}
\caption{ Markov chain representation of the configuration dynamics. When the configuration is $\mn$, the dissociation rate of a cluster of size $i$ into a cluster of size $k$ and one of size $i-k$ is $k_bm_i$, while
the rate of formation of a cluster of size $i+j$ from two clusters of size $i$ and $j$ is equal to $\frac{k_f}{2}m_im_j$. }
\label{figm1mN}
\end{center}
\end{figure}

\begin{figure}[ht!]
\begin{center}
\includegraphics[scale=0.55]{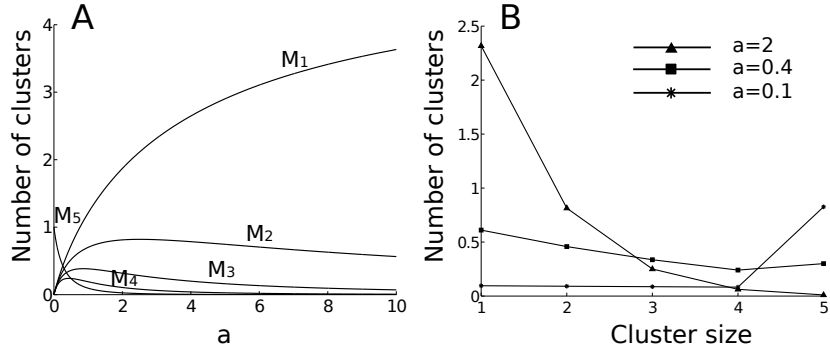}
\caption{Mean number $<M_n>$ of clusters of size $n$ as a function of the parameter $a$ (A) for $N=5$ particles (equation \eqref{meansize}) and (B) mean number of cluster for $a=0.1$, $a=0.4$ and $a=2$ as a function of the cluster size.}
\label{G1}
\end{center}
\end{figure}

This consideration defines the steady state. Indeed, if we consider
the equilibrium probability distributions associated with the Markov
chain configuration $\mn$, then the transition between the two
neighboring states $(m_1,\dots,m_i-1,\dots,m_j-1,
\dots,m_{i+j}+1,\dots,m_N)$ and $\mn$, is obtained first from the
coagulation of a cluster of size $i$ with one of size $j$, with
a rate $C(i,j) = \frac{k_f}{2}m_im_j$ if $i\neq j $ and
$C(i,i)=\frac{k_f}{2}m_i(m_i-1)$ otherwise. The factor
$\frac{1}{2}$ accounts for the two cases $C(i,j)$ and $C(j,i)$. The
fragmentation rate from $i+j$ to $(i,j)$ is $F(i,j) = k_b
(m_{i+j}+1)$ (Fig. \ref {figm1mN}). Thus, the stationary probability $\pi$ necessarily
satisfies the relation
\beq
 \frac{\pi(m_1, \dots,m_i-1, \dots,m_j-1, \dots,m_{i+j}+1, \dots,m_N)}{\pi \mn} = \frac{C(i,j)}{F(i,j)}=
 \frac{1}{2a}\frac{m_im_j}{m_{i+j}+1}.
 \label{probabilitypi}
 \eeq
When the configuration $\mn$ is made of $K$ clusters, the
configuration $(m_1, \dots,m_i-1, \dots,m_j-1, \dots,m_{i+j}+1,
\dots,m_N)$ has $K-1$ clusters. From Eq. (5), we have that
$$ \Pi_{K-1}(a) = \Pi_K(a) \frac{K(K-1)}{2(N-K+1)a}$$
and the conditioned stationary probability satisfies
\beq
 \frac{\pi(m_1, \dots,m_i-1, \dots,m_j-1, \dots,m_{i+j}+1, \dots,m_N)| K-1)}{\pi(m_1,\dots,m_N| K)} &=& \frac{ \Pi_K(a)}{ \Pi_{K-1}(a)} \frac{1}{2a}\frac{m_im_j}{m_{i+j}} \\
&=&\frac{N-K+1}{K(K-1)}\frac{m_im_j}{m_{i+j}+1}. \label{durett}
\eeq
A direct computation shows that the probability $p\mn$ satisfies
equation \eqref{durett} \cite{Durrett}.
{Now using Bayes rule, the joint probability of the configuration
$\mn$ and $K$ clusters is the product of the conditional probability
$p(m_1,\dots,m_N |K)$ by the probability of having $K$ clusters}
\beq
p(m_1,\dots,m_N,K )=p(m_1,\dots,m_N |K)\Pi_K.
\eeq
We shall now estimate the size of clusters from the above
considerations. For a total of $N$ particles, when there are $K$
clusters, the mean number of clusters of size $n$ is
\beqq
<M_n>_K&=&\sum_{(m_i)\in \ANK'}m_np(m_i|K)  \\
     &=& \frac{(N-n-1)!K!(N-K)!}{(N-1)!(K-2)!(N-n-K+1)!},
 \eeqq
which we computed using equation \eqref{explicit} and that when a cluster of size $n$ is contained in the distribution $\mn$, it is equivalent to have $N-n$ particles in $K-1$ clusters. Finally the mean number of clusters of size $n$ is obtained by summing over all possibilities when there are $K$ clusters,
\beq
 <M_n>&=&\sum_{K=1}^N <M_n>_K\Pi_K\\
 &=& \frac{(N-n-1)!}{(N-1)!} \sum_K \frac{K(K-1)(N-K)!}{(N-n-K+1)!} \Pi_K. \nonumber
 \label{Mia}
 \eeq
Using the expression for $\Pi_K$ \eqref{eq:Pik}, we obtain
\beq
 \left \lbrace
\begin{array}{ccc}
<M_n>&=&2a\ds{\frac{{_1F_1}(-N+1+n;2;-2a)}{{_1F_1}(-N+1;2;-2a)}} \mbox{ if } n<N,\\
&&\\
 <M_N>&=&\ds{\frac{1}{{_1F_1}(-N+1;2;-2a)}}.\\
\end{array}\right.
\label{meansize}
\eeq
In addition, the mean number of clusters of size $n$ is $2a\frac{\Pi_{1}(N)}{\Pi_{1}(N-n)}$, \textit{i.e.} this is the probability ratio of having one cluster when there are $N$ particles over the probability of having one cluster when there are $N-n$ particles.
Finally, the variance of the number of clusters of size $n$ is, if $N>2n+1$,
\beq
<V_n> &=& <M_n>-<M_n>^2 +2a \frac{N-2n-1}{(N-2n+1)(N-2n+2)} \times\bigg( \frac{\Pi_1(N)}{\Pi_1(N-2n)}-\frac{\Pi_1(N)}{\Pi_1(N-2n-1) } \bigg)
\eeq
and otherwise
 \beq
<V_n> &=& <M_n>-<M_n>^2.
\eeq
To summarize this analysis, we plot in Fig. \ref{G1} the mean number of clusters of size $n$ for five particles. We will also use this analysis in the final section to study telomere clustering in yeast.\\
At equilibrium, particles are exchanged between clusters. To
characterize the exchange of telomeres between clusters, we now
estimate the probability $p_2(N,a)$ to have two given particles in
the same cluster $C$ for a fixed number of particles $N$ and
equilibrium constant $a$. For a distribution of clusters
$(n_1,\dots,n_K)$, the probability for two specific particles to be
in $C$ is obtained by choosing the first particle in the cluster
$n_i$, which is equal to the number of particles in the cluster
divided by the total number of particles $\frac{n_i}{N}$ and thus
the probability to have the second particle in the same cluster is
$\frac{n_i-1}{N-1}$. Summing over all clusters, we get
\beq \label{eqq}
P(2 \hbox{ particles } \in C | (n_1,..n_K))&=&\sum_{i=1}^K
\frac{n_i}{N}
\frac{n_i-1}{N-1}\\
&=&\frac{1}{N(N-1)}(\sum_{i=1}^Kn^2_i-N).\nonumber
\eeq
Using that $\sum_{j=1}^{N} n_j^2 = \sum_{i=1}^Ni^2m_i$, we have
\beq
\sum_{(n_j)\in
\ANK}p(n_j)\sum_{j=1}^{K}n^2_j=N+2N\frac{N-K}{K+1}.
\eeq
Summing now over all distributions of clusters, the probability
\beq
p_2(N,a)=-\frac{2}{N-1} + \frac{N+1}{N-1}\frac{_1F_1(-N+1;3;-2a)}{_1F_1(-N+1;2;-2a)}.
\label{proba2telomeres}
\eeq
Surprisingly, the three-term recurrence relation for Kummer's function (\cite{AbraSteg} Eqs. 13.4.1-13.4.6) gives
$ _1F_1(-N+1;3;-2a)=   \frac{N-1}{N+1}{_1F_1}(-N+2;3;-2a)  + \frac{2}{N+1}{_1F_1}(-N+1;2;-2a).$
Finally
\beq
p_2(N,a) =  G_1(a,N).
\label{P2model2}
\eeq
\begin{figure}[ht!]
\begin{center}
\includegraphics[scale=0.7]{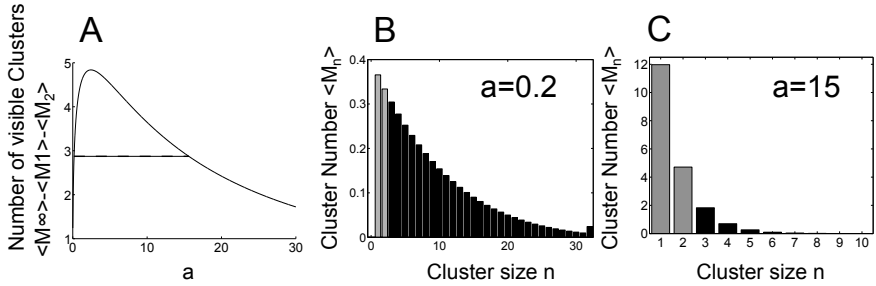}
\caption{(A) Number of clusters containing more than three particles $O(N,a)$ as a function of the equilibrium parameter $a$, for $N=32$. Experimentally, 2.9  clusters were observed in average, which corresponds to $a=0.2$ and $a=15$. (B, C) Distribution of the mean number of clusters as a function of the cluster size (Eq. \eqref{meansize}), for $N=32$ and $a=0.2$ (B) and $a=15$ (C). Grey bars indicate single particles and clusters of two particles, which are not observable.  }
\label{figexp}
\end{center}
\end{figure}
\section{Application to telomere clustering in yeast}
To apply the previous analysis to the organization of telomere
clustering in yeast, we use a coarse-grained model of a telomere,
the motion of which can be approximated as Brownian \cite{Shavtal}.
In addition, because clusters are of small sizes, the arrival time
of a telomere to cluster is Poissonnian. Indeed, it is the time for
a stochastic particle moving on the nuclear surface to reach a small
target. Our goal is now to show that telomere clustering can be
described by our diffusion-fragmentation-association model and thus
by the Master equation \eqref{Markov1}. For that purpose, we propose
to estimate the forward and backward rates, constrained by recent
observations that in yeast the 32 telomeres form 2 to 8 clusters,
with an average of 2.9 clusters per cell, containing in average 4
telomeres \cite{Gotta,Ruault}.

For a telomere radius $r=0.015$ $\mu$m, moving on a sphere (nucleus
surface) of radius $R=1$  $\mu$m with a diffusion coefficient
$D=0.005$ $\mu$m$^2$.s$^{-1}$ \cite{Bystricky}, we find using a
Brownian simulation for two telomeres to meet, that the forward rate
is $k_f=1.9 10^{-3}$ s$^{-1}$. Although the cluster can vary in size
when telomeres attach, we shall make the assumption that the
encounter will stay constant and use this rate in our previous
Markov modeling. Indeed, telomeres are attached to the nuclear
surface by a family of proteins Sir2/Sir3/Sir4 \cite{Gotta,Ruault}.
Telomere clusters are elicited by the formation of Sir3-Sir3
interactions, and Sir3 abundance is directly related to the
stability of the clusters (dissociation rate $k_b$). Thus, in the
absence of any experimental evidences, we consider that telomere
diffusion is driven by a molecular complex, moving on the nuclear
surface. When this complex has the shape of a cylinder, the
diffusion constant is given by the log of the radius \cite{saffman}.
Since clusters contain at most four telomeres \cite{Ruault}, the
effective radius does not change much leading to a small change of
the diffusion constant. This justifies our constant approximation
for the encounter rate $k_f(m,n)$ of two clusters of size $n$ and $m$.

The dissociation rate $k_b$ of a telomere from a cluster cannot be
easily derived from experimental data because clusters made of one
or two telomeres are not experimentally visible \cite{Ruault}. Using
the results of the first sections (equations
\eqref{Minf},\eqref{meansize}), we use our model Markov model to compute
and to plot (Fig.
\ref{figexp}A) the number of visible clusters obtained by the
formula
\beqq
O(N,a)&=& <M_{\infty}(a,N)>-<M_{1}(a,N)>-<M_{2}(a,N)>\nonumber\\&=& 1-4a+(a(N-1)+6a^2)G_1(a,N) - \frac{4}{3}a^3G_2(a,N)
\eeqq
as a function of the equilibrium parameter $a$. We find from this
graph that there are two possible values for $a$, giving an average
of 2.9 visible clusters, which are $a=0.2$ and $a=15$. To select the
correct value, we built the cluster distributions associated for
these two values from equations \eqref{meansize}. We further
obtained the mean number of clusters of a given size (Fig.
\ref{figexp}B,C). However, for $a=0.2$, telomeres form a giant cluster,
containing almost all the telomeres, which is not reported
experimentally. Finally, we conclude that $a=15$, for which there
are always less than five telomeres per cluster, in agreement with
experimental observations \cite{Ruault}. Furthermore, we obtain for
the dissociation rate, the value $k_b=a k_f=0.03$ s$^{-1}$, which
was not known before. Finally, using formula \eqref{P2model2}, we
can predict that two specific telomeres are located in the same cluster approximately 4\% of the time.\\

In conclusion, we obtained here a novel analysis for a
coagulation-fragmentation process of a finite number of Poissonnian
particles restricted in a bounded domain. We used our model to
estimate the dissociation rate $k_b$ of telomeres from a cluster, a
constant that can reveal the local organization of telomeres in a
cluster. Although we applied this analysis to the organization of
telomeres in yeast, it could be applied to study telomere
organization for general organisms.





\bibliographystyle{elsarticle-num}







\end{document}